\documentclass[letterpaper, 10 pt, conference]{ieeeconf}  

\IEEEoverridecommandlockouts                              

\title{\LARGE \bf
HapKnob - A Motorized Shape-changing Haptic Knob Interface}

\author{Zhili Gong$^{1}$, Zitong Wei$^{1}$, and Jeremy D. Brown$^{1}$
\thanks{$^{1}$ Department of Mechanical Engineering, Johns Hopkins University, Baltimore, MD 21218, USA. Email: {\tt\small \{zgong13, zwei32,jdelainebrown\} @jh.edu} }%
}
\begin{document}

\maketitle
\thispagestyle{empty}
\pagestyle{empty}

\begin{abstract}

The absence of physical interfaces creates challenges when interacting with touchscreen technology. This study aims to investigate an innovative haptic solution for interacting with graphical user interfaces. A motorized shape-changing rotary knob interface, HapKnob, has been developed, achieving seven distinctive shape configurations and various force feedback renderings. HapKnob presents a compact design and can provide additional configurations and combinations based on user needs. The anticipated results hold the potential to advance the development of future user interfaces, especially in situations where visual interaction is unavailable.

\end{abstract}

\section{INTRODUCTION}

In modern society, traditional physical interfaces, such as buttons, are being gradually replaced by touch screens, resulting in users losing the tactile feedback associated with manipulating physical devices. This shift underscores the need to reevaluate current control methodologies and develop creative solutions to alleviate the visual burden required for task completion. Recognizing this demand, the objective of this work is to investigate a novel haptic interface capable of conveying information through modular shape changes and dynamically rendered force feedback.

Various shape-changing interface prototypes have been developed over the last 20 years with different purposes and functionalities, such as expressing emotions, rendering textures, and enabling distanced physical interaction \cite{c1,c2}. Likewise, force feedback has been used in many applications, including teleoperation and medical simulation/training. Force feedback has also been used to improve graphical user interfaces \cite{c3}.

Knobs function as ubiquitous physical controllers in our daily routines, enabling users to directly input parameters into a system. Given the widespread utility and existence of knobs, their design and potential for modification have motivated this study. Prior research and commercial product development have delved into the viability of crafting shape-changing knobs, integrating force feedback into knob functionality, and the combination of both elements \cite{c4,c5}. Yet, only a limited number of them have effectively replicated the size commonly encountered in daily knob usage. These demonstrations have also been limited to a maximum of five shape features and visual-dominant task scenarios. 

\begin{figure}[t]
      \centering
      \includegraphics[scale = 0.145]{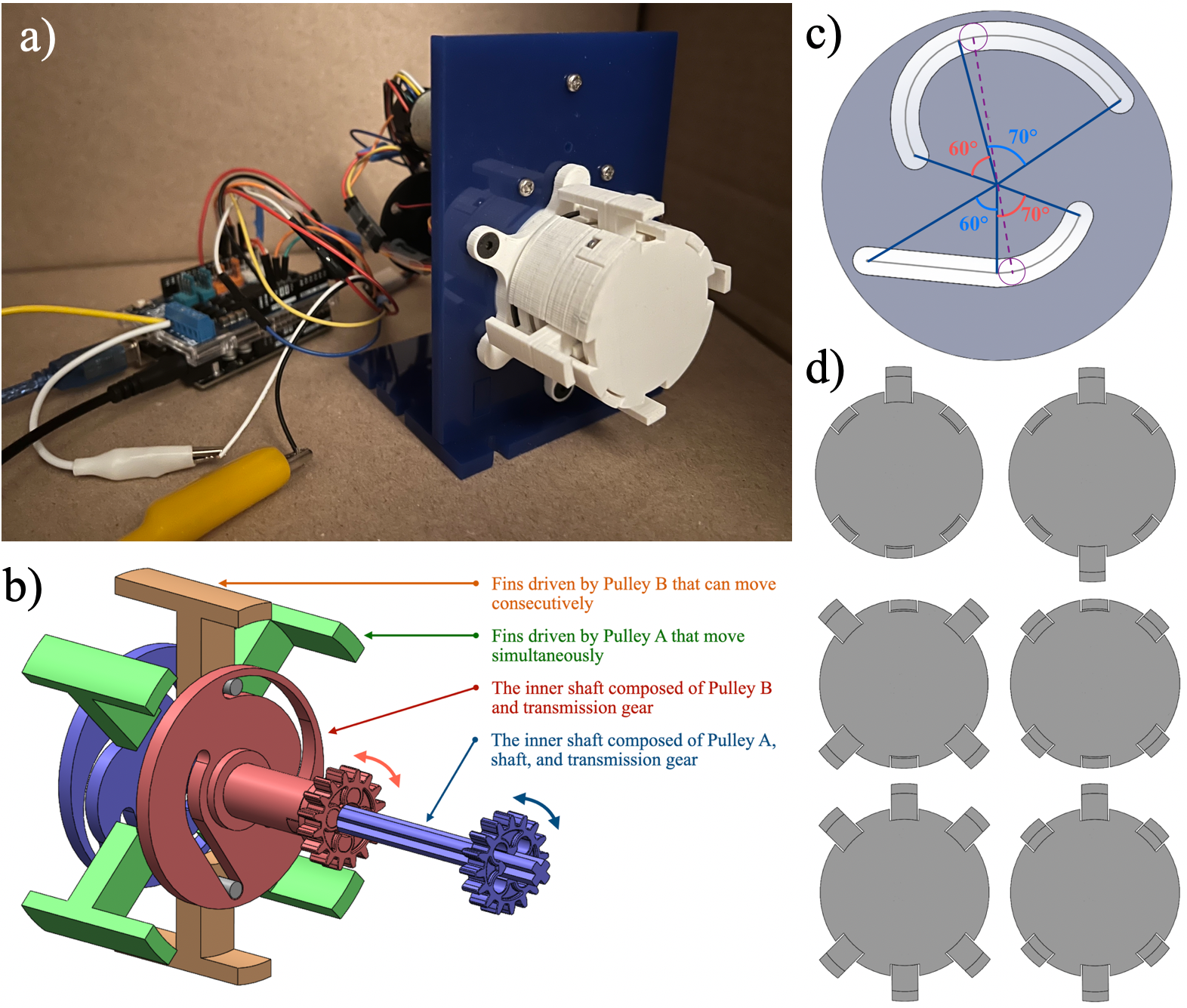}
      \caption{a) The proposed motorized shape-changing haptic knob, HapKnob, exhibits 4 fully expanded fins driven by Pulley A; b) Breakout view of the coaxial expansion pulleys mechanism; c) Geometry of slots cut on Pulley B, with purple geometries indicating the locking position; d) Demonstration of the 6 unique shape configurations that HapKnob can achieve.}
      \label{Fig1}
\end{figure}

This research aims to address these gaps by creating a shape-changing knob that emulates the typical size of everyday knobs while broadening the range of attainable shape features. HapKnob, a motorized, shape-changing knob as seen in Fig.~\ref{Fig1}.a, is developed with the objective of improving task performance under limited vision. This innovative approach seeks to extend the applicability of shape-changing knobs beyond visually dominant contexts, contributing to a more versatile and user-friendly interface design.

\section{METHODS}

HapKnob is a shape-changing haptic knob interface designed to deliver both force feedback and shape sensations. The dynamic shape-changing capability is attained through six servo-controlled fins, while variable force feedback is realized using a motor. The knob assembly of the HapKnob has a diameter of 52\,mm and a height of 30\,mm, mimicking the size of knobs encountered daily. 

\subsection{Shape-Changing Mechanism}

The shape-changing feature is achieved by two coaxial expansion pulleys driving six fins situated in rails as seen in Fig.~\ref{Fig1}.b. Pulley geometries are designed so that the fins can be linearly displaced by 8\,mm in the radial direction. The outer diameter of the fin matches the diameter of the knob frame, creating a continuous smooth side wall when the fins are fully contracted. Pulley A and Pulley B drive 4 fins and 2 fins, respectively, and their rotation can be individually controlled due to the coaxial shaft design. Pulley A drives all four fins simultaneously, and they all displace the same amount when the pulley rotates. Pulley B also drives two fins simultaneously, but the two fins move consecutively due to special slot geometry Fig.~\ref{Fig1}.c. A 10-degree surplus is designed on pulley B slots for the fin to naturally lock in place under the pointer mode, where only one fin is expanded as shown in the upper left corner of Fig.~\ref{Fig1}.d. The geometry of the slots on pulleys also determines how much force is needed to move the fin at different positions. Current geometries are selected because a relatively low torque is required to move the fins by rotating the pulleys, while a greater amount of force is required to move the pulleys by pushing or pulling the connected fins. The pulleys are driven by two SG90 servos at a 1:1 ratio, using two 16-tooth gears rotating 90 and 130 degrees, respectively. The servos and the shape-changing knob are stationary relative to each other once assembled. A slip ring is used to separate the rotation of the wires on the knob assembly and the base structure. 

\subsection{Actuation for Force Feedback}

The HapKnob utilizes a 12\,V C25GA370 DC Motor, featuring a 1:9.7 speed-reduction gearbox and an integrated quadrature encoder (24 CPR), to generate torque output as the system's force feedback. A 24-teeth drive gear is press-fitted to the gearbox output shaft to transmit power to the knob via the ring gear integrated with the shape-changing unit. Through the incorporation of the gearbox and the 24:60 (2:5) drivetrain, the system achieves a resolution of 0.6186 degrees per count at the knob. Dynamic torque is generated based on the knob position and the designed rendering feature. Although the current powertrain setup introduces impedance to the system that is usually sub-optimal in haptic devices, friction and damping experienced during rotation are considered to be close to that of a knob in real applications.

\subsection{Control}

The Arduino UNO, with an attached Arduino motor shield, controls both the motor and servos. The encoder provides a signal that, following a sequence of computations, determines the knob's position relative to its initial point. The positional data acquired undergoes differentiation and filtering to retrieve the rotational speed. Subsequently, force can be applied corresponding to the knob's status, with negative values for counterclockwise torque, and positive values for clockwise torque. The control algorithm of HapKit, an open-sourced haptic toolkit \cite{c6}, is utilized to compute the required PWM signal based on the specified torque.

\section{PRELIMINARY RESULTS}

\begin{figure}[thpb]
      \centering
      \includegraphics[scale = 0.27]{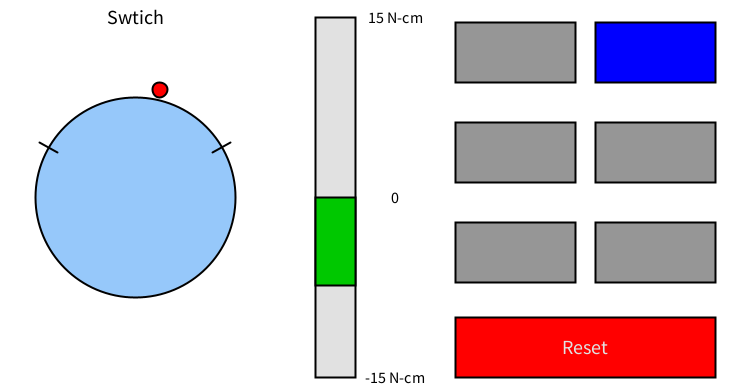}
      \caption{GUI built to monitor the system status. The red dot moves along the blue circle to display the knob's position, the green bar indicates torque intensity, and the button array enables users to set and reset modes.}
      \label{Fig2}
\end{figure}

Seven unique shape configurations have been designed and tested to be easily distinguishable by first-time users. In addition to the default configuration, where all fins are contracted to form a cylindrical knob, the remaining six configurations are demonstrated in Fig.~\ref{Fig1}.d. Notably, the potential for creating more configurations exists by expanding fins with finer increments. Various fundamental force feedback features have also been successfully created, including hard walls, bumps, valleys, linear damping, and textures, as well as a variable mass-spring-damper system. Furthermore, a simple graphical user interface has been built for demonstration purposes using Processing software \cite{c7}. This interface allows users to track their operations and choose between six different modes, each offering distinct force rendering and shape configurations as seen in Fig.~\ref{Fig2}. Mechanically, under the current hardware setup, the maximum torque output can reach over 25\,N-cm without significant system instability, and a maximum of 20 bumps or valleys (spaced 18 degrees apart) can be noticeably rendered.

\section{FUTURE WORK}

The existing low-end gearbox presents noteworthy backlash, introducing uncertainties in position tracking that subsequently contribute to system instability in specific force feedback renderings. To address this issue and improve user experience, the existing drivetrain will be substituted with one that is backlash-free, enabling more precise and consistent position tracking.

Despite the utility of the existing GUI, future development involves improving the graphical user interface to enhance engagement and overall user experience. Additionally, GUIs play a vital role in guiding participants through upcoming user study experiments, specifically evaluating HapKnob's performance in tasks designed for non-visual conditions. 

Identifying potential applications for this design concept is also imperative. One potential application involves integrating HapKnob as a universal tangible object within VR or AR devices, thereby expanding its utility and relevance in immersive environments.

\end{document}